\documentstyle[epsfig,12pt]{article}

\begin{document}
\begin{center}

\large

{\bf Giant vesicles at the prolate--oblate transition:\\ A macroscopic bistable
system}


\large \vspace{0.5cm} {\sl
Hans-G\"{u}nther D\"{o}bereiner $^{\ast\ \dagger\ }$ and \
 Udo Seifert $^{\ast}$} \vspace{0.5cm} \normalsize

$\rm{^{\ast}}$Max-Planck-Institut f\"{u}r Kolloid-
und Grenzfl\"{a}chenforschung,\\
Kantstra\ss e 55, 14513 Teltow-Seehof, Germany\\
$\rm{^{\dagger}}$Department of Physics, University of British Columbia,\\
 6224 Agriculture Road, Vancouver, British Columbia, Canada V6T 2A6\\

\vspace{0.5cm}
\vspace {1cm}
\end{center}

\def\beq{\begin{equation}}
\def\ee{\end{equation}}

\begin{abstract}
  Giant phospholipid vesicles are shown to exhibit
thermally activated transitions
  between a prolate and an oblate shape on a time scale of several
  seconds.   From the fluctuating contour of such a vesicle we
  extract ellipticity as an effective reaction coordinate whose
  temporal probability distribution is bimodal. We then reconstruct
  the effective potential from which we derive an activation energy of
  the order of $k_BT$ in agreement with theoretical calculations.  The
  dynamics of this transition is well described within a Kramers model
  of overdamped diffusion in a bistable potential. Thus, this system can serve
as a model for  macroscopic
  bistability.\\ \mbox{}
\end{abstract}

\eject

Macroscopic bistable systems for which thermal energy is sufficient to
overcome the barrier in observable times are rare, quite in contrast to
microscopic bistability so ubiquitous in nature \cite{haen90}.
In this paper, we consider bistability in the shape space of giant ($10 \mu m$)
phospholipid vesicles  \cite{lipo91b}.
The shapes of these vesicles are determined by the minima of the
 bending energy of
the vesicle membrane subject to constraints on the total area and enclosed
volume \cite{canh70,helf73,evan74}.
 A cut of the energy landscape along a path
connecting nearby local minima then defines
  an effective double well potential in shape space.
Since this  potential is created by the vesicle  due to its bending elasticity,
our system is qualitatively different from the most
prominent example of macroscopic bistability studied so far which is
 a one micron silica bead caught in a double well potential generated
by an optical trap \cite{z:simo92}. In the latter
 system, the (optical) potential is
created externally  in real space via gradients of the electromagnetic field.

A vesicle (shape) is not a static entity but shows thermally excited
fluctuations, which are
clearly visible by video microscopy, due to the softness of the vesicle
membrane.
 At constant external parameters such as temperature, these fluctuations
 occur around a well-defined  mean shape corresponding to the global
minimum of bending energy.
As temperature changes, these mean shapes usually evolve smoothly except
at  phase boundaries denoting so-called discontinuous transitions
 along which shapes of different symmetry have
the same energy.  In this report, we discuss the observation of vesicles at
the prolate-oblate transition \cite{hein93,z:jari95}. These vesicles exhibit
thermally activated
jumps between the two states at constant external parameters.
In Fig.~1, we show the contours of  a
typical sequence of vesicle shapes at equidistant times (6.3 s). A
vesicle is seen to change its shape from a prolate with the mean symmetry axis within the focal plane to an oblate with the  mean symmetry axis perpendicular
to the focal plane and vice versa \cite{foot1}.

\begin{figure}
\begin{center}
  \leavevmode
  \epsfig{file=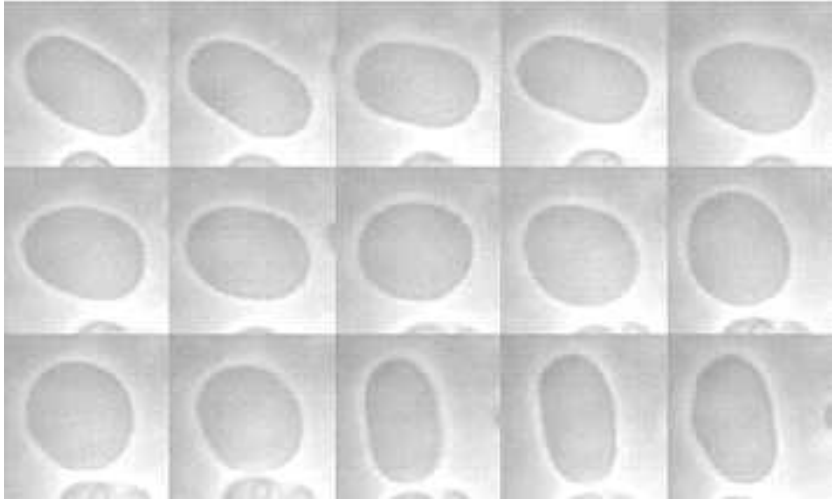,width=11cm}
\end{center}
\caption{
Prolate-oblate fluctuations of a DMPC vesicle. The snapshots are taken
at equidistant time intervals of $\Delta t = 6.3\ s$. The
vesicle is seen to change its shape from a prolate with the mean symmetry axis
within the focal plane to an oblate with the  mean symmetry axis perpendicular
to the focal plane and vice versa.
Vesicles were prepared from stearoyl-oleoyl-phosphatidylcholine (SOPC) or
dimyristoyl-phosphatidylcholine (DMPC) in sucrose solution (50 mOsm) and
incubated in excess glucose solution (48 mOsm). They were
observed by
video phase-contrast microscopy in a microchamber at constant temperature.  Due
to the slightly higher density of sucrose, the vesicle sink to the bottom of
the observation chamber and rest gently against the glass wall. Thus, the free
rotational diffusion of vesicle shapes is restricted to the 2 dimensions of the
focal plane of the microscope. This stabilisation procedure is crucial for
recording time series of vesicle shapes. Suitable vesicles were selected and
observed near the boundary between the prolate and the oblate phase.}
\label{sequence}
\end{figure}

For a quantitative analysis, the contours of the vesicles
in the focal plane are determined
by image analysis \cite{fauc89,duwe90,z:doeb95a}.
We describe the shapes using  polar coordinates,
\begin{equation}
  \label{expansion}
  r(\phi)\ =\ r_0\left(1 + \sum_{n\geq2}a_n \cos(n \phi)
+ \sum_{n\geq2} b_n \sin(n \phi)\right)\ ,
\end{equation}
where $a_1=b_1=0$ implies the definition of the center of gravity of each
contour as the origin.
Since gravity holds the vesicles on the bottom of the chamber, oblate shapes
have a circular contour while prolate shapes (with their major axes
parallel to the bottom) look ellipsoidal \cite{z:krau95}. Thus, it is
 sufficient to monitor the amplitude $A \equiv \sqrt{(a_2^2+b_2^2)} > 0$,
which effectively measures the ellipticity of the vesicle contours.

\begin{figure}
\begin{center}
  \leavevmode
 \epsfig{file=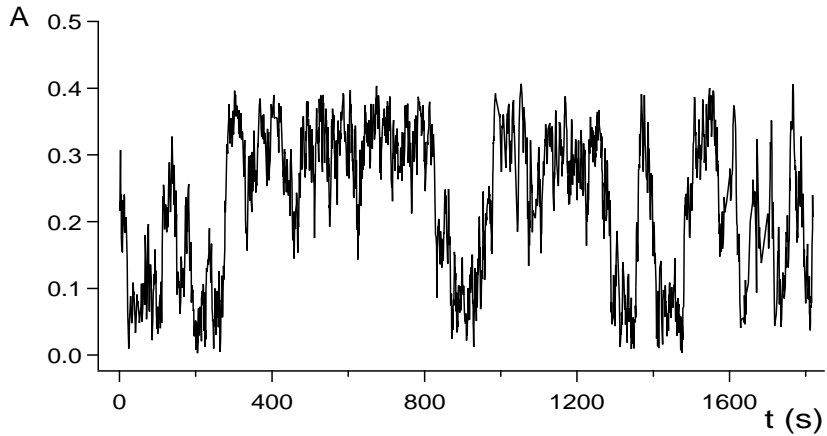,width=12cm}
\end{center}
\caption{
Time series of vesicle ellipticity $A=\protect\sqrt{a_2^2+b_2^2}$ near the
prolate
oblate
transition. The ellipticity exhibits fluctuations around two distinct mean
values
which correspond to the prolate and
oblate shape respectively. The shapes are quantified by extracting the vesicle
 contour via image analysis. Briefly, a contour point is defined by the
gradient
 of the phase contrast profile across the membrane with respect to the local
background level. Thereby, one finds a closed loop made up of points with
sub-pixel
resolution. The amplitudes $a_2$ and $b_2$ are then obtained from this contour
via Eq.~(\protect\ref{expansion}).
}
\label{timeseries}
\end{figure}

 In Fig.~\ref{timeseries}, a time series of $A$ for a typical vesicle is shown.
 The apparent jumps in  $A$ between a small numerical value near zero and a
larger positive number reflect the bistable nature of the system, already
evident
in Fig.~\ref{sequence}. The larger values correspond to the fluctuating prolate
shape, whereas the small values parallel the oblate vesicle with its
fluctuating equatorial circular contour. 

\begin{figure}
\begin{center}
  \leavevmode
 \epsfig{file=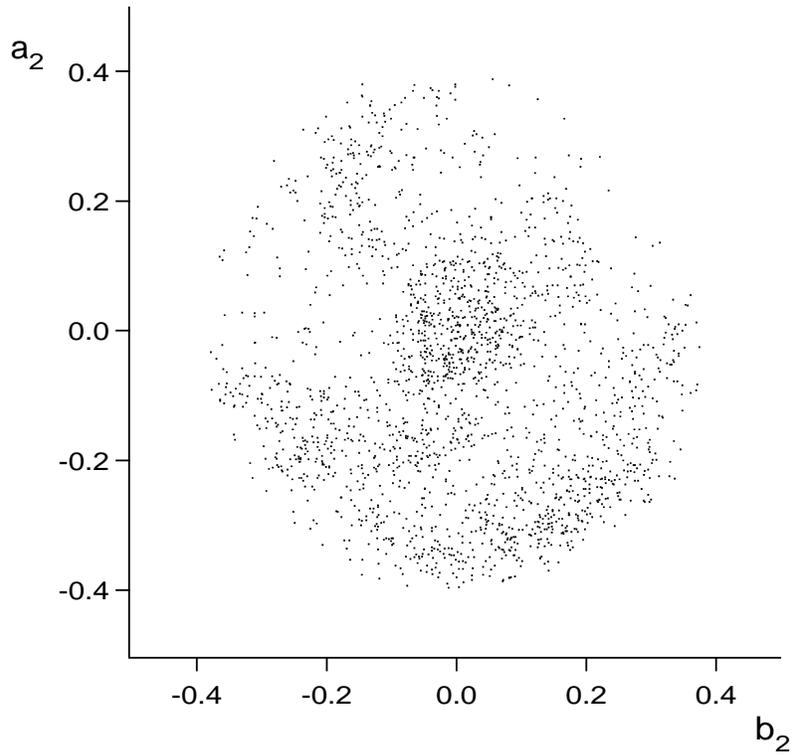,width=12cm}
\end{center}
\caption{
Distribution of instantaneous contour parameters $(a_2,b_2)$ in a polar
diagram. The dark center corresponds to the oblate shape, the darker
rim to the prolate phase. Since the activation energy is of the order
of $k_BT$, there is significant population also along the saddle points.}
\label{shapespace}
\end{figure}

The diffusion in shape space is evident in
Fig.~\ref{shapespace}, where we show the distribution of
 $(a_2,b_2)$. 
The points  cluster at the center
(the
oblate phase) and along a circle (the prolates) because of the rotational
symmetry of the problem.

\begin{figure}
\begin{center}
  \leavevmode
 \epsfig{file=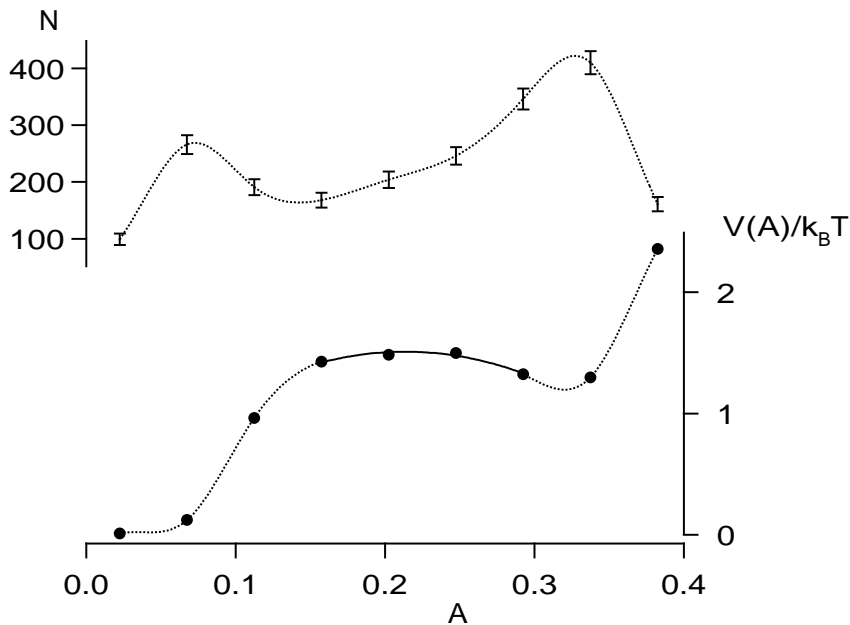,width=12cm}
\end{center}
\caption{
Histogram of events $A$ of the ellipticity obtained with a
bin size 0.045. The dotted lines are guides to the eye. The double well
 potential $V(A)$ is then reconstructed by using Eq.~(\protect\ref{PotEqu}).
 The
saddle point location at $A_s=0.21$ and its curvature $w_s^2=29\ k_BT$ are
obtained
via a fit (solid line) of the saddle to a parabola.}
\label{potential}
\end{figure}

We now construct an effective potential for the variable $A$ from the
measured probability distribution $P(A)$, represented by a histogram of $A$ as shown in Fig.~\ref{potential}.
The probability
to find the shape in the interval  $[A,A+dA]$
(and any orientation $\alpha$) is proportional to  $dA \ A e^{-V(A)/k_BT}$,
where $V(A)$ is
an effective potential. The additional
factor $A$ arises from changing from cartesian coordinates $a_2,b_2$ to
polar coordinates $A$ and $\alpha$. From the measured  $P(A)$, we thus get
\beq
\label{PotEqu}
V(A)/k_BT = \ln A -\ln P(A) + const
\ee
as shown in Fig.~\ref{potential}. The activation energies $Q_{ob,pro}$ from the
oblate and the prolate phase, respectively,  thus obtained,
are $Q_{ob}\simeq 1.5\pm 0.1\ k_BT$ and $Q_{pro}\simeq 0.2\pm 0.1\ k_BT$.

Although the prolate shape is  metastable with a small barrier height
$Q_{pro}$, it is nevertheless populated with a higher probability than
the oblate minimum (compare Fig.~\ref{timeseries}). This is due to the
 measure factor $A dA$ for the probability distribution $P(A)$, which
 reflects the larger density of states (with energy $V(A)$) of the
prolate compared to the oblate shapes. All prolate vesicles with a
fixed ellipticity $A$ but varying orientation $\alpha$  have the same
energy $V(A)$, while the oblates lack this orientational degree of
freedom.

This observation of a vesicle as fluctuating in a bistable potential  \cite{foot2}
is qualitatively different from the three other fluctuation effects observed in
vesicles so far. (i) Fluctuations of quasispherical
vesicles have long been used to extract the bending rigidity of bilayer
membranes\cite{fauc89,duwe90,evan90,z:nigg95}. Our experiment shows that this
technique must fail if the
available excess area of the vesicles is so large that the discontinuous
character of the prolate-oblate transition becomes relevant.
(ii) The budding transition \cite{evan90,kaes91,miao94}, where a small satellite is
expelled from a parent
vesicle,
 is a strong first-order transition for which the activation energy cannot be
surmounted thermally. One then observes pronounced
fluctuations as the spinodal is approached \cite{z:doeb95a}.
(iii) Conformal diffusion of vesicles of higher topology involves diffusion
 along a one parameter continuous curve in a ``flat'' potential
 \cite{juel93b,z:mich95}.

A quantitative comparison of our experimental results
 with theoretical first-principle calculations is delicate.
First, the presence of gravity makes even a
calculation of the shape of lowest energy non-trivial \cite{z:krau95}.
However, since we
expect gravity to be a minor correction of the energy, one can,
in a first
approximation,  compare with the energy of free vesicles. Ignoring all
shape fluctuations,  this energy
has been calculated for the two local minima (prolate and oblate) as well
as for the intermediate stationary saddle point which is a non-axisymmetric
ellipsoid \cite{hein93,z:jari95}.
One finds a weak first-order transition with a
maximal energy barrier between prolate
and oblate for reasonable material parameters of the order
$Q^0_{ob,pro}\simeq 1\ k_BT$. In principle, this energy barrier acquires
an additional entropic
 contribution  from the available phase space for fluctuations orthogonal to
the
reaction co-ordinate $A$. While for axisymmetric
shapes, these entropic
corrections can be calculated, a calculation for the
non-axisymmetric
saddle shape is far beyond presently available techniques. Since the
measured activation energies are in reasonable agreement with the theoretical
value, our
results indicate that these entropic corrections do not differ strongly for
the three stationary shapes involved.

\def \V{V_{\rm eff}}
The dynamical aspects of this transition can be modeled as an overdamped
Kramers escape problem in a  potential  $V(a_2,b_2)=V(A)$
\cite{kram40,z:gard94}.
The
diffusion
in the angle co-ordinate can be integrated out, which leads to an
one-dimensional diffusion process for the variable $A$ in the
effective potential $\V(A)\equiv V(A) -k_BT \ln A$. The corresponding Langevin
equation reads
$
\gamma \partial_t A = - \partial \V/\partial A + \zeta
$,
where $\zeta$ is a stochastic force with the usual correlations
 $\left<\zeta (t)\zeta(t')\right> = 2k_BT\gamma \delta (t-t')$. The
effective friction coefficient $\gamma$ is  assumed to be
independent of the shape and given by $\gamma =const \ \eta R^3$, where $\eta$
is the viscosity of water and $R$ the apparent radius of the oblate. The
Kramers mean escape time from the oblate minimum becomes \cite{z:gard94}
\beq
\tau_{ob} = {\gamma \sqrt{2\pi k_BT}\over \omega_{ob}^2 \omega_s A_s}
\exp{Q_{ob}/T} ,
\label{eq:rate}
\ee
where the frequencies $\omega_{ob,s}$ are defined by a quadratic approximation
to the energy in the oblate ($V(A) \approx V(0) + {\omega_{ob}^2\over 2} A^2$)
and at the saddle at $A_s$ ($V(A) \approx V(A_s) - {\omega_s^2\over 2}
(A-A_s)^2$).

The pre-exponential factor in (\ref{eq:rate})
 becomes more transparent if it is expressed as
 $t_{ob}\sqrt{2\pi k_BT}/\omega_s A_s\ $ in terms of measurable quantities.
Here,  $t_{ob}\equiv \gamma/\omega_{ob}^2 $ is the
relaxation
time for $a_2$ or $b_2$ fluctuations in the oblate phase. The
 measured dynamical correlation
functions   $\left<a_2(t)a_2(0)\right>$ and $\left<b_2(t)b_2(0)\right>$
 can both be fitted against simple relaxational behavior with
a decay time  $t_{ob}\simeq  5\pm 2\  s$. Using
the measured values $Q_{ob}/k_BT\simeq 1.5\ $, $A_s\simeq 0.21$ and
$\omega_s^2\simeq 29\ k_BT$ as extracted from the
potential (see Fig.~\ref{potential}), we obtain  $\tau_{ob}\simeq 50 \pm 25$ s.
This
value is in reasonable agreement with the mean residence time $\bar \tau_{ob}
\simeq
76 \pm 8 $ s
 in the oblate phase as extracted directly from the time
series $A(t)$ as shown in Fig.~\ref{timeseries}.

We now turn to a discussion of the prolate phase. For the fluctuations of the
 ellipticity  $A$ around its mean value $\left<A\right>$, we find a
relaxation time $\tau_A \simeq  8\pm 1  s$.
Fluctuations in $\alpha$ amount to rotational diffusion.
For the latter, we find a rotational diffusion coefficient
$ D = (5.7\pm 0.2) \times 10^{-3}$ rad$^2$/s.  It is this free diffusion which
is
responsible for the relative stability of the metastable prolate state.
The mean escape time from the prolate phase should not
be written in a Kramers form since the implicit assumption of
Kramers theory that the activation energy is larger than $k_BT$
  is violated on the prolate side.

To summarize, we have reported the observation of a macroscopic bistable system
as evidenced by
large shape fluctuations of giant vesicles between a prolate and oblate state.
This result confirms
the theoretical prediction of a weak discontinuous transition between prolate
and
oblate shapes. The mean
escape time out of the metastable oblate state was found to be in reasonable
agreement with the
value obtained within a Kramers approach for the effective radial
potential.  Thus, fluctuating vesicle shapes may serve as a model system for
further
 studies of macroscopically slow
 diffusion processes in a bistable potential.

We are grateful for discussions with E. Evans, M. Jari\'{c}, M. Kraus, R.
Lipowsky, and \mbox{M. Wortis}.


\begin{thebibliography}{10}
\bibitem{foot1}
For a prolate shape, the mean symmetry axis is its major axis, whereas for an
oblate shape,  the mean symmetry axis coincides with the minor axis.
\bibitem{foot2}
We use the term ``bistable'' even though one of the locally stable shapes, the prolate,
has a trivial one-fold degeneracy due to its orientational degree of freedom.
\end{thebibliography}
\end{document}